\documentstyle[apjpt4]{article} % APj
\singlespace
\begin{document}
\twocolumn 

\title{Morphology and kinematics of Planetary Nebulae.
I.~A new modeling tool.}

\author{C. Morisset\altaffilmark{1,2}, R. Gruenwald\altaffilmark{2},
\and S. M. Viegas\altaffilmark{2} } 

\altaffiltext{1}{Laboratoire d'Astronomie Spatiale, Traverse du Siphon, Les 
Trois Lucs, 13012 Marseille, France $<$morisset@astrsp-mrs.fr$>$}
\altaffiltext{2}{Instituto Astron\^omico e Geof\'{\i}sico, USP,
Avenida Miguel Stefano, 4200 CEP 04301-904 S\~ao Paulo, SP, Brazil
$<$ruth@agn1.iagusp.usp.br$>$ and $<$viegas@agn1.iagusp.usp.br$>$}

\begin{abstract}

We present a new modeling tool for planetary nebulae, 
based on 3D photoionization calculations. Our goal is to show 
that all the information provided by observations,
regarding kinematics and morphology, have to be
consistently  accounted for, in order to get a real insight of the object.
Only 3D simulations offer this possibility. From models for two theoretical 
PNe, we show that the  enhancement
in the  equatorial zone
observed in several PNe is not necessarily due to a density gradient, as
usually interpreted. It is also shown that asymmetric velocity
profiles often observed (e.g., Gesicki et al. 1998) can  be easily reproduced.
Observations providing a better insight on the
morphology of the PN are discussed. 

\end{abstract}


\section{Introduction} 

Planetary Nebulae (PNe) show different morphologies, but since the
works of Kwok et al. (1978) and \cite{B87}, 
it is possible to reproduce their observed shape basically with two
types of geometry: spherico-elliptical ones and bi-polar ones
(also called butterfly), all of them axi-symmetrical in a first
approximation. 

Observed morphological differences may result 
from (1) different orientation
of the line-of-sight and of the axis of
symmetry and (2) the age of the nebula. Morphological sequences have been
studied by Balick (1987), Pascoli (1990) and Zhang \& Kwok (1998),
among others.
Evolutionary theories based on the generalized interacting stellar wind model
(GISW, see  Franck \& Mellema 1994 and references therein) seem
to confirm these interpretations. 

Information obtained from photometric imaging and intermediate resolution 
spectra have been complemented by narrow band images (e.g., 
Sahai et al. 1997) and high resolution spectra (e.g., Walton et al. 1990), 
providing information on the kinematics of the nebula. 

The various observations can not be correctly interpreted without
detailed modeling, which must consistently reproduce all observational
data. For some years, most of the data have been analysed by GISW models
(Soker \& Livio 1989; Mellema, Eulderink \& Icke 1991; Icke, Balick
\& Frank 1992), which account only for the gas dynamics. 
A more realistic model to analyse morphologies and kinematics of PNe
was proposed by Frank \& Mellema (1994). These 2D simulations include
radiation processes and ionization balance, while the gas cooling
through line emission is obtained from polynomial approximations
for the line emissivities (Balick et al. 1993). 

Here a new modeling tool for studying
the morphology and kinematics of PNe is presented. 
Assuming a velocity field, the results of a 3D photoionization code 
(Gruenwald, Viegas \& Brogui\`ere 1997, hereinafter GVB) 
are used to generate images 
to be directly confronted with various kinds of observations.
We address general PNe issues,
raised by the observations, as gas distribution and morphology,
brightness enhancement at the equatorial ring, the asymmetric
emission line profiles (e.g. Gesicki et al. 1998).

The computational methods used to model planetary nebulae
are presented in \S 2.  Results for two 
theoretical objects are shown and discussed in \S 3. 
The conclusions appear in \S 4.

\section{Computational methods}
\subsection{The 3D photoionization code}\label{sub-sec-3d}

The 3D photoionization code is described in GVB.
The input parameters are the ionizing radiation spectrum,
the gas chemical composition and the spatial density
distribution. The nebula is divided in a great number of cells. 
Inside each cell the physical conditions, which are obtained from the 
ionization equilibrium and thermal balance, are assumed homogeneous.
For each cell, the output are the emissivity of atomic lines and  
the physical conditions of the gas
(electronic density, fractional abundances, electronic temperature).
The number of cells defines the numerical accuracy. It depends on 
the specific object under analysis, as well as on the computer memory.

The effect of the diffuse radiation can be accounted for. However, 
in its present version, 
the required computational time is too high (few tens of the on-the-spot 
case). Thus, we assume here the on-the-spot approximation, using case B 
parameters of P\'equignot et al. (1991) for the hydrogen
recombination coefficient.
The effect is small enough and do not affect the general conclusions 
presented here. 
The H$\beta$ luminosity is underestimated by 30$\%$ or less
when the on-the-spot approximation is assumed.  The percentage depends on
the stellar temperature: the higher the stellar temperature the bigger the
underestimation  of the H$\beta$ luminosity. 
The differences between the line intensities relative to H$\beta$ 
is less than 20$\%$, 
except for the weak lines ($<$ 0.05 H$\beta$).

Emission line images are obtained after a line-of-sight integration.
The calculations are made assuming a stationary approximation.
In order to analyse the kinematics, a velocity field is assumed after
obtaining the photoionization results for each cell.
The data cubes, containing the results generated by the 3D code for 
the cells, are 
read by an IDL (RSI) code which provides the tools for filtering,
rotation, projection, statistics, velocity maps and echellograms, 
as described below.

In this paper, only axi-symmetric nebulae will be studied. 
In order to increase the number of cells used in the calculations
(increasing the numerical accuracy), 1/8 of the nebula is modeled.
For the calculated 1/8 of the nebula, the number of cells is 
$50^3 = 125\ 000$.
In this case, the ionizing source is placed at one of
the corners of the large cube.
In order to restored the whole nebula, with the ionizing source at the center,
the remaining 7/8 are added up, 
using the calculated 1/8  
after appropriate rotations.

\subsection{Rotations}

The nebula is 
defined in a XYZ coordinate system, where X is the axis of
symmetry, the origin being at the
ionizing source (center of the rebuilt cube). The rotations and projections
operate on a data cube of 
$xyz$ coordinates. In the following, the line of sight used for the
projections will be $y$.
Before any rotation the origins, as well as the axes of the 
XYZ and $xyz$ coordinate systems, are coincident.

The rotations of the data cubes are performed in the $xyz$ space, 
plane by plane, using the two-dimensional 
ROT function of IDL, in a $\sqrt{3}$ times greater cube to avoid losing
information. If needed, up to three rotations around the three axes of the
cube can be performed, in any order.

If rotations around more than one axis are necessary, 
it is easier if one starts with a rotation 
around an axis perpendicular to both the axis of symmetry of the 
nebula ($X$) and 
the axis along the line-of-sight ($y$). 
In order to compare the theoretical images to observations, a  
second rotation of the data cube may be needed. 

\subsection{Emission line images}
\label{sub-images}

For each emission line the 3D code provides the luminosity
in each cell. The total luminosity (corresponding to the whole nebula),
as well as the intensity corresponding to a given aperture,
can be obtained and compared to observational data.

The projected image on the plane of the sky is 
obtained for a given emission line
by integrating the emissivities of the cells lying along the 
line-of-sight which is taken perperdicular
to one of the cube faces.

Maps of emission line intensity ratios can also be obtained, 
especially those used to obtain the PN physical conditions.
These maps can be transformed 
into density or temperature maps using relationships 
(e.g., Alexander \& Balick 1997)
between the projected line intensity ratios and these quantities.

\subsection{Line profiles}

In order to obtain a line profile, 
a cube of velocities at the same resolution as the 3D code output is
generated according to a predefined law. 
In this paper we assume a radial velocity field.
Since the projection is made on the $y$-axis direction,
only the corresponding component
of the velocity is retained ($V_y(x,y,z)$). For each emission line,
a velocity profile is generated, for each line-of-sight, by 

$$ 
\phi_\lambda (v,x,z) = \sum_y {{\epsilon_\lambda (x,y,z)} \over
{\sqrt{\pi}. \xi(x,y,z)}} . e^{ - {[{{\Delta V(v,x,y,z)} \over
{\xi(x,y,z)}}]}^2}  
$$ 

\noindent
with:
$$
\Delta V(v,x,y,z) =  V_y(x,y,z) - v
$$
$$
\xi(x,y,z) = \sqrt{V_{th}^2(x,y,z) + V_T^2 }
$$
$$
V_{th}(x,y,z) = \sqrt{2kT_e(x,y,z)/Am_H}
$$

\noindent
where $\epsilon_\lambda (x,y,z)$ is the emissivity of a given 
line inside the $xyz$ cell, $V_T$ is the turbulent velocity 
and $V_{th}(x,y,z)$ is the thermal velocity of the corresponding atom of mass
number A. The local electronic temperature $T_e(x,y,z)$ is
consistently obtained from the 3D code.

Velocity contour maps and P-V  maps (echellograms) can be generated. 
After masking through an aperture and convoluting with an
instrumental profile, the theoretical data can be compared 
to observations. 

\subsection{Average physical quantities}
\label{sub-stat}

For each emission line, average physical quantities of the region where 
the line is produced can be obtained. Average values for the electron 
temperature, density, hydrogen ionization, and velocity are calculated by 
weighting theses quantities  by the line emissivity.
The general assumed formula is:
$$
<Q>_\lambda = {{\sum Q(x,y,z).\epsilon_\lambda (x,y,z)}
\over {\sum \epsilon_\lambda (x,y,z)}}
$$
where $Q$ is replaced by $T_e$, $n_e$, $H^+/H$ or $v$. These quantities can be
obtained for the whole nebula, as well as for lines-of-sight characterized
by an $xz$ position, summing on $xyz$ or $y$, respectively.
Notice that Peimbert (1967) defined some of these 
quantities weighting by $n_e.n(X^i)$ instead of by the line emissivity.

Dimensionless temperature fluctuations ($t^2$) and  density
fluctuations ($n_e^2$) can also be computed: 
$$
Q^2_\lambda = {{\sum
(Q(x,y,z)-<Q>_\lambda)^2.\epsilon_\lambda (x,y,z)} 
\over {<Q>_\lambda^2 \sum \epsilon_\lambda (x,y,z)}}
$$

Empirical methods for chemical abundance determination are 
usually based on these averaged quantities and fluctuations, 
defined for the whole nebula. They have been generally applied 
to point-to-point observations, assuming a spherical symmetry. 
In this case, however, the method 
is not always correct, even for spherically symmetric objects (Gruenwald 
\& Viegas 1992).
A real nebulae is not necessarily symmetric and homogeneous. Therefore, 
only 3D models can provide tools for improving the empirical methods.

\section{Application to planetary nebulae}

\subsection{Models}

Two simple theoretical PNe will be assumed (A and B) with two possible
geometries (1 and 2). Models A correspond to higher black-body temperature,
higher density and higher luminosity than models B (Table \ref{tab-spectro}).  
On the other hand, type 1 models have a
spherical cavity surrounded by a shell with a density 
gradient; the density increases linearly with the angle, from the 
pole to the equator.
Type 2 models have a prolate elliptical cavity surrounded 
by a constant density shell. 
These two geometries are two simple cases which can reproduce the 
bipolar brigthness enhancement usually shown by PN images. Surely, 
the geometry of real objects is
determinated by hydrodynamics. Thus, a further improvement, out of
the scope of this paper, would be to use the 3D photoionization code
adopting a geometry obtained from a hydrodynamic code.

The input parameters related to the
density distribution are given in Table \ref{tab-spectro}.  
The following notation is used: n$_{H}$(cav) is the cavity density; 
n$_{H}$(shell) is 
the shell density (type 2 models), or its range (type 1);  R$_{inner}$
is the cavity radius (type 1) or the semi-axis of the elliptical 
cavity (type 2). 
Models A are inspired on the canonical PN of the 
Paris workshop (P\'equignot 1986, see also Ferland et al. 1995).

The adopted chemical abundances are the same for the four models, 
and are taken from the canonical PN, i.e., H:1.0, He:0.1, C:3(-4), N:1(-4), 
O:6(-4), Ne:1.5(-4), Mg:3(-5), Si:3(-5), S:1.5(-5), Cl:4(-7), Ar:6(-6),
Fe:1(-7). 

For all models, the PN is radiation bound in all directions.

\subsection{Spectroscopical results}

Results for the 3D models A1, A2, B1, and B2, corresponding 
to the whole nebula, are listed in Table \ref{tab-spectro}.
Only line intensities of the  main emission lines, relative 
to H$\beta$, are given, as some average quantities obtained from line
ratios, using Alexander \& Balick (1997) fits or the 
formulae given in \S \ref{sub-stat} (lower part of the table). 

The differences between the results of A1 and A2 models are not 
significant. 
For both models, the results for line intensities 
are within the values obtained 
for the canonical PN (spherical symmetry and constant density) with
the various 1D photoionizing codes (Ferland et al. 1995).
Notice that our results were obtained using the on-the-spot 
aproximation (see \S \ref{sub-sec-3d}). From the point of view of the 
emission-line spectra, as well as of the
derived physical
conditions, no conclusion about the nebula geometry can be drawn, 
since both geometries lead to similar spectroscopical results.
The same is true for a planetary nebulae with a lower stellar temperature,
lower luminosity and density, as seen from the similarity between the 
results of models B1 and B2.

For both (A and B) types of nebula, the
determination of temperature and density, from the line ratios 
or from the formulae (\S \ref{sub-stat}), are
consistent within 10$\%$. 
The electronic temperatures obtained from the [OIII] and [NII] 
line ratios are different because these lines come from different regions.
\placetable{tab-spectro}

\subsection{Imaging}

Emission line images often show a brightness enhancement which is
usually interpreted as the signature of a density gradient in
the equatorial belt of PNe (\cite{W68}; Khan \& West 1985;
Aaquist \& Kwok 1996; \cite{ZK98}). 
The images of  H$\beta$, [O III]$\lambda$5007, He II$\lambda$4686, and 
[N II]$\lambda$6583 obtained from models A and B are given, respectively, 
in Figs. \ref{fig-AA} and \ref{fig-BB}. The nebulae are projected
along the Y axis (no rotation).
The top four panels correspond to type 1 models, while the botton four ones 
to type 2 models. Notice that type 1 and 2 models have different geometries,
which might reproduce the observed enhancement. In type 1 models, the gas
density increases linearly with the angle from the pole to the equator, and 
is distributed around an inner spherical cavity. On the other hand, in type 2 
models the gas is uniformly distributed around a prolate elliptical cavity.
As suggested by the figures, the models are able to generate the brightness 
enhancement. 
Notice that the line intensity enhancement is due to the higher density 
in the equator for type 1 models, whereas, for type 2 models, it 
appears because the density of the {\em ionizing photons} reaching
the inner edge of the gas is higher in the direction along 
the minor axis.

However, the real geometry can not be easily inferred from the shape 
suggested by the images.  In fact, model A1 images show an ellipsoidal 
shape, while for model A2 the shape suggests a spherical distribution,
which is the opposite of the adopted shapes for the inner cavities.
This contrasting behavior is less pronounced in models B, 
with lower stellar temperature and luminosity, as well as with 
lower density, than models A. The image shape given by model A2 is
defined by the ionizing radiation dilution, independently of the gas 
distribution. On the other hand, in model B2 the main factor defining 
the image shape is the shape of the inner cavity.
 
From these results, we see that narrow band imaging may not 
differentiate the nebular geometry as previously assumed. 
In the next section we discuss a possible geometry indicator.
\placefigure{fig-AA}
\placefigure{fig-BB}

\subsection{Density maps}
\label{sub-dens-map}
As shown above, spectroscopic results and emission line imaging 
do not provide enough contraints on the nebula density distribution. 
The density indicator generally used is the [SII] line intensity ratio.
The density maps obtained from the [S II]
ratio  (see \S \ref{sub-images}) are 
shown in Fig. \ref{fig-rsii-A}
for models B1 (top panel)  and B2 (bottom panel). 
As expected, the density gradient adopted for model B1 is
clearly perceived.
\placefigure{fig-rsii-A}

\subsection{Velocity profiles}
High resolution observation of PNe emission lines often 
show double-peaked profiles, interpreted as the emission of a 
geometrically thin expanding shell. Usually, the two
peaks are not symmetric (see Gesicki et al. 1996, 1998 for
seven PNe, and \cite{DWW99} for NGC3132). If the asymmetry were due to 
local extinction, the blue peak should always be more intense than the 
red one. Since 
this is not the case, another explanation for the asymmetry must be
found. 

In the following, the effect on the line profiles, due to geometry and
orientation, will be discussed. We show that the asymmetry in the 
line profiles can be reproduced using a 3D photoionization code and 
the numerical tools described above (\S 2). In order to illustrate 
these results, one case 
is presented, using model A1 with a given velocity law.
A linear law, as used by Sahu \& Desai (1986), is quite natural. In the 
following 
we adopted $\vec V = \alpha\times \vec r/r_o$, with $\alpha$ = 20 km/s 
and $r_o$ = 2.4$\times$10$^{17}$cm (maximum of the [O~III]
emissivity). 
The velocity profiles obtained after 
the rotation of the nebula by 50$^o$ around the $z$-axis and
projection along the $y$ axis are 
shown in Fig. \ref{fig-velo_A}. 
Each profile corresponds to a synthetic observation at a different 
position in the nebula. The assumed positions and aperture 
are shown in the upper panel, over
the [O III] image. A turbulent velocity of 2 km/s was taken into
account. Such value for the turbulent velocity is small, so that 
the adopted velocity field dominates the line profile.
No convolution through an instrumental profile was applied. 

As seen in the figure, the profile changes from one region to 
another. Depending on the position, asymmetric profiles are obtained, 
including asymmetric double-peaked profiles. 
For positions at the right side of the nebula
the double peaked profiles, when asymmetric, 
show the blue peak more intense than 
the red one. On the other hand, due to the rotation around the 
$z$-axis, a red asymmetry (not shown in the
figure) would be obtained for positions on the left side of the
image. 

The mean velocity (as defined in \S \ref{sub-stat}) of the [O III] zone is 
$<V>_{5007}$ = 22.1 km/s, which is
comparable to what could be found from the peak separation
on the profile corresponding to the synthetic observation of
the central part of the nebula.
\placefigure{fig-velo_A}

Other velocity laws, suggested by hydrodynamic
calculations, could be adopted when discussing observations from
real objects.

\subsection{PV diagrams}
More accurate than line profiles observed at discrete positions, 
Position-Velocity (PV) diagrams (also called echellograms) 
provide an effective diagnostic for the
geometry of PNe. In the case of line tilts (Guerrero et al. 1997),
PV-diagrams give constraints
on the inclination of the PN (\cite{M98}).
The observations and the 2D model of NGC 650-1 (\cite{B96}) is another 
example of the use of PV-diagrams to deduce the geometrical properties
of PNe. Such a model is not detailed enough to reproduce
the differences between the observed emission of [N II]$\lambda$6584 and 
[O III]$\lambda$5007.
A 3D photoionization code is needed.
PV-diagrams obtained from our model A1 rotated by
50$^o$ around the $z$ axis and projected on the $y$ direction are shown 
in Fig. \ref{fig-ech_A}. 
The 
upper-left panel shows the [O III]$\lambda$5007 image of the nebula.
The upper-right
and lower-left panels show PV-diagrams obtained through vertical and 
horizontal 
centered slits, respectively. The lower-right panel illustrates 
iso-velocity contours, for V = 9 km/s.
The $z$-PV-diagram shows the signature of the tilt
of the PN around the $z$-axis, while the absence of an asymmetry in the 
$x$-PV diagram indicates that the Z axis of the nebula is perpendicular 
to the line of sight and coincides with the $z$-axis. 
Such simulation is available from the 3D code
for all the emission lines and is a very powerful tool to model high 
resolution observations like the ones from the TAURUS instrument 
(Walton et al. 1990).

\placefigure{fig-ech_A}

\section{Concluding remarks}

This paper illustrates the importance of 3D photoionization simulations
when analyzing the various observational data currently obtained for PNe.
Regarding the stellar temperature and luminosity, two different PNe are 
simulated. For each one, two density distributions 
are assumed, in order to test the constraints on the nebula 
geometry provided by the data.   
For each kind of observation (emission line spectroscopy,
emission line imaging, emission line profiles) 
the results obtained with the two different geometries are compared.

Regarding line emission from the whole nebula, the emission line ratios
do not discriminate between the two possible density distributions
since the two models have the same average density, giving similar 
[S II] line intensity ratios. Emission line imaging showing 
an ellipsoidal shape with an  intensity enhancement on the minor axis
is obtained with both density distributions. Thus, this kind of
observational data do not provide a method to distinguish between the 
two distributions. Density maps, obtained from emission line ratios
that are density indicators, can provide one of the keys to 
the geometry puzzle, as shown in \S \ref{sub-dens-map}. The other key, related
to the orientation of the axis of symmetry of the nebula,  comes from
the emission line profiles observed in different locations of the nebula.
An asymmetric double-peaked profile is usually observed in PNe, but until
now never theoretically reproduced by photoionization models. It can be 
generated assuming that the 
expansion
velocity increases outward and that the axis of symmetry of the nebula
makes an angle with the plane of the sky (see \S 4). 
The axis of symmetry tilt can also be visualized on the echelograms (\S 5).
Notice that double-peaked profiles are also generated by pure hydrodynamic 
models (Mellema 1994).

In brief, the different kinds of observations that are now available for
planetary nebulae can only provide a complete insight of the object
and its properties, if they are confronted with the results of 
photoionization simulations from a 3D code.

\acknowledgements
C. Morisset is thankful to IAGUSP for the hospitality during his 
post-doctoral visit.
This work was supported by CNPq (n.~304077\-/\-77-1 and
n.~150162\-/\-96-0), FAPESP  
(n.~98\-/\-01922-7), and PRONEX/FINEP (n.~41.\-96.\-0908.\-00). 
The calculations were performed 
with a DEC\-/\-ALPHA 3000 workstation (FAPESP n. 93\-/\-4348-6).

\newpage
\figcaption[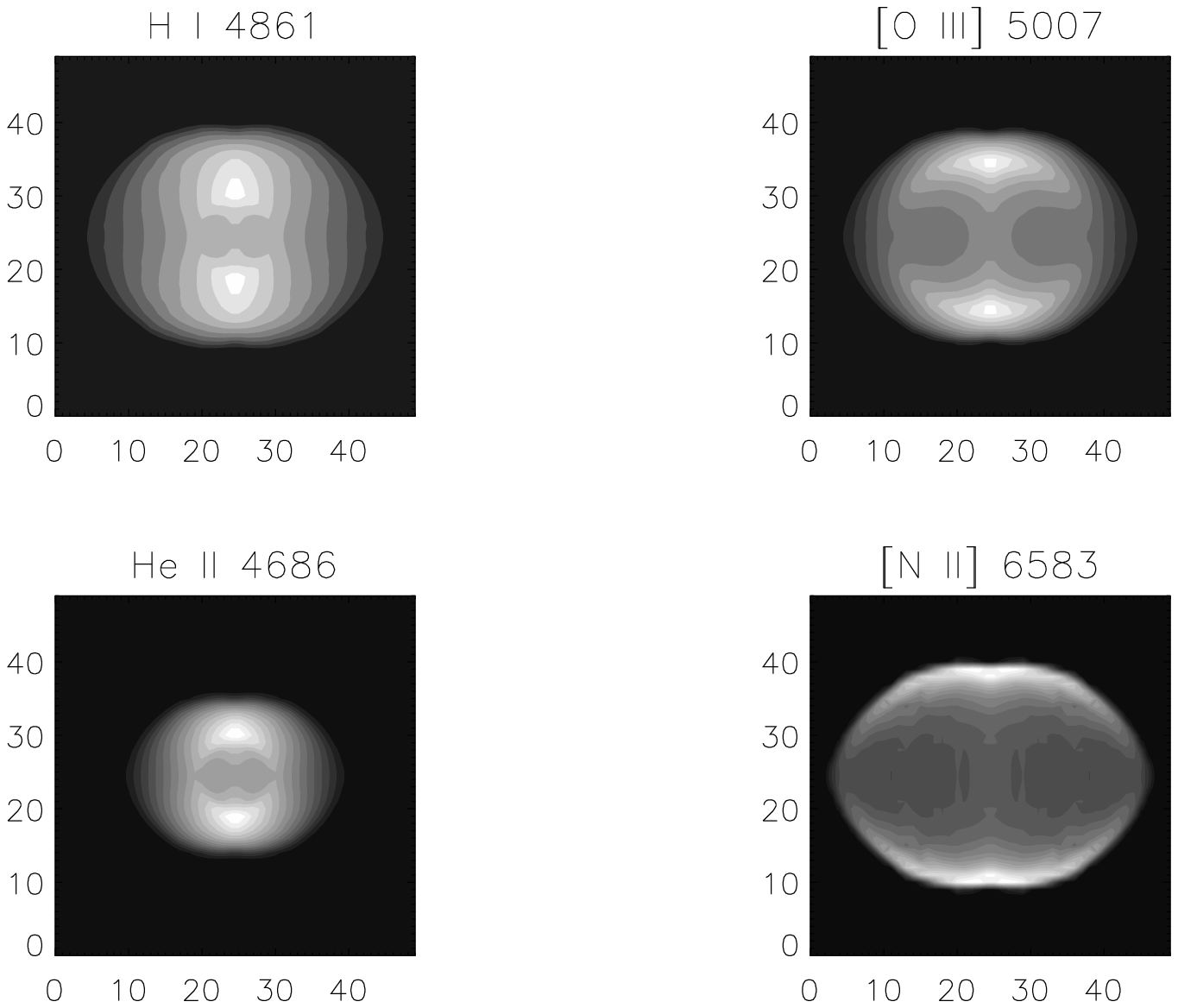,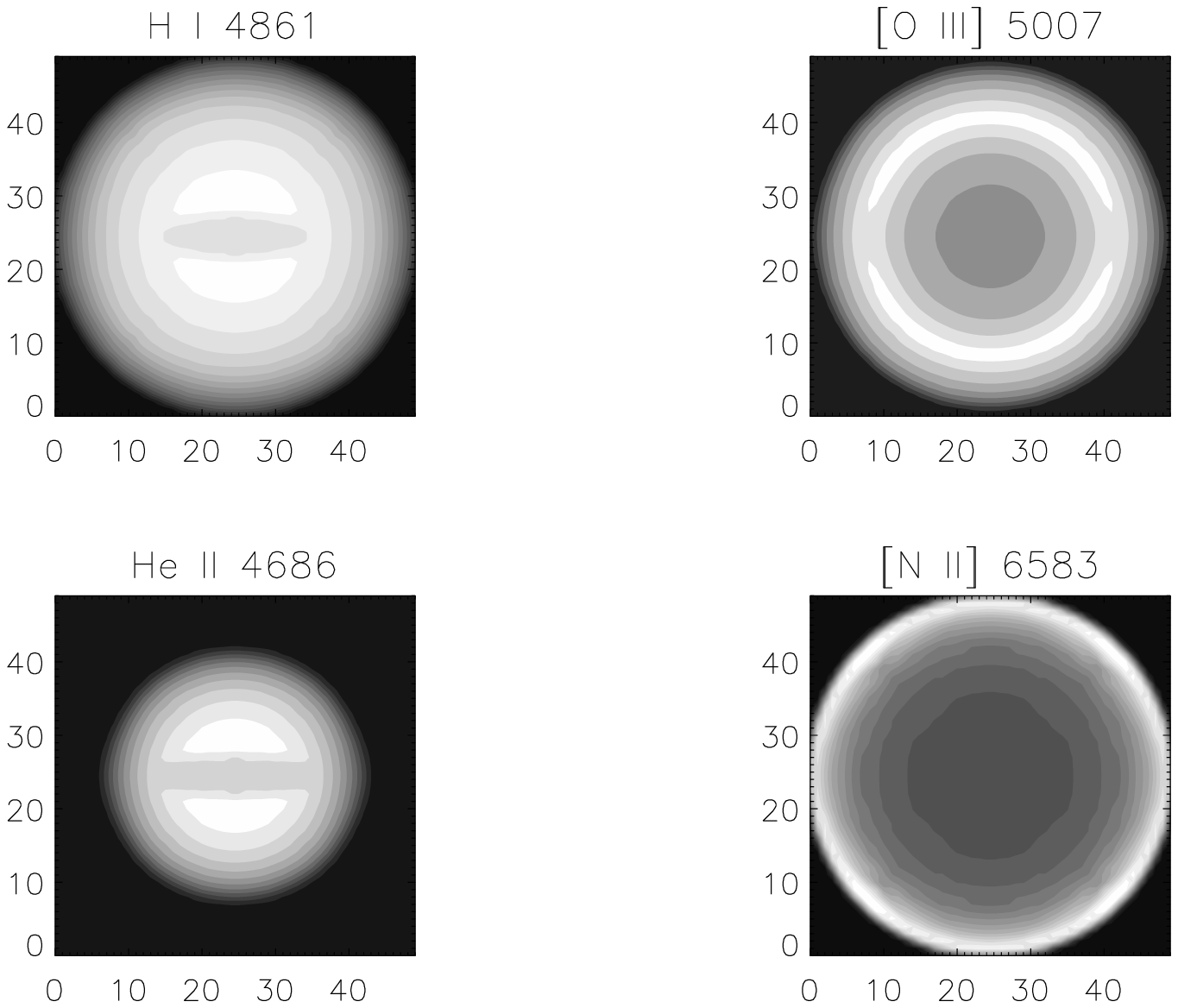]{
Emission-line imaging from models A1 (top four panels) and 
A2 (botton four panels). For each model, H$\beta$, [O III]$\lambda$5007,
He II$\lambda$4686, and [N II]$\lambda$6583 intensity maps are shown. Both 
axes are in pixels.
\label{fig-AA}}

\figcaption[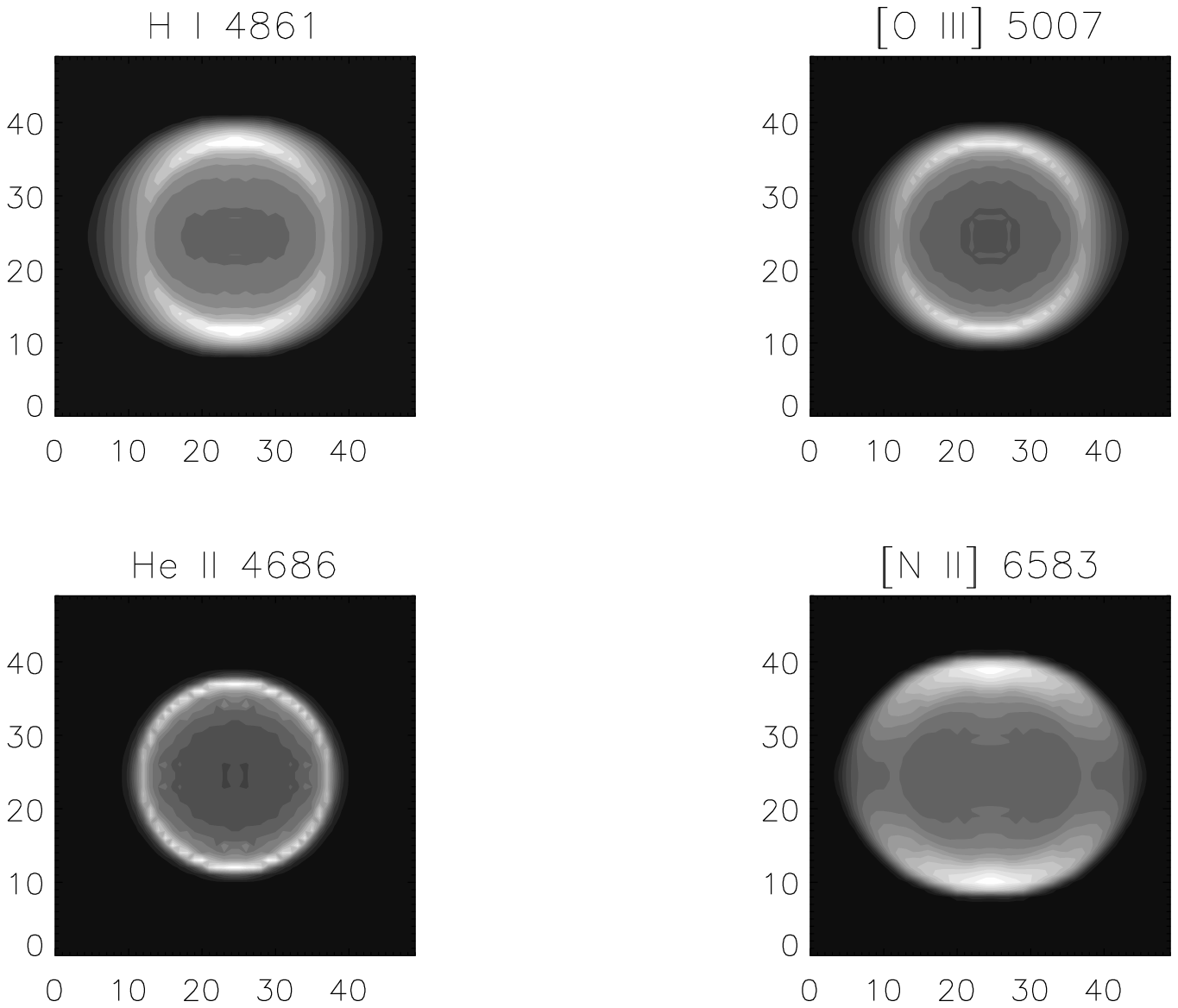,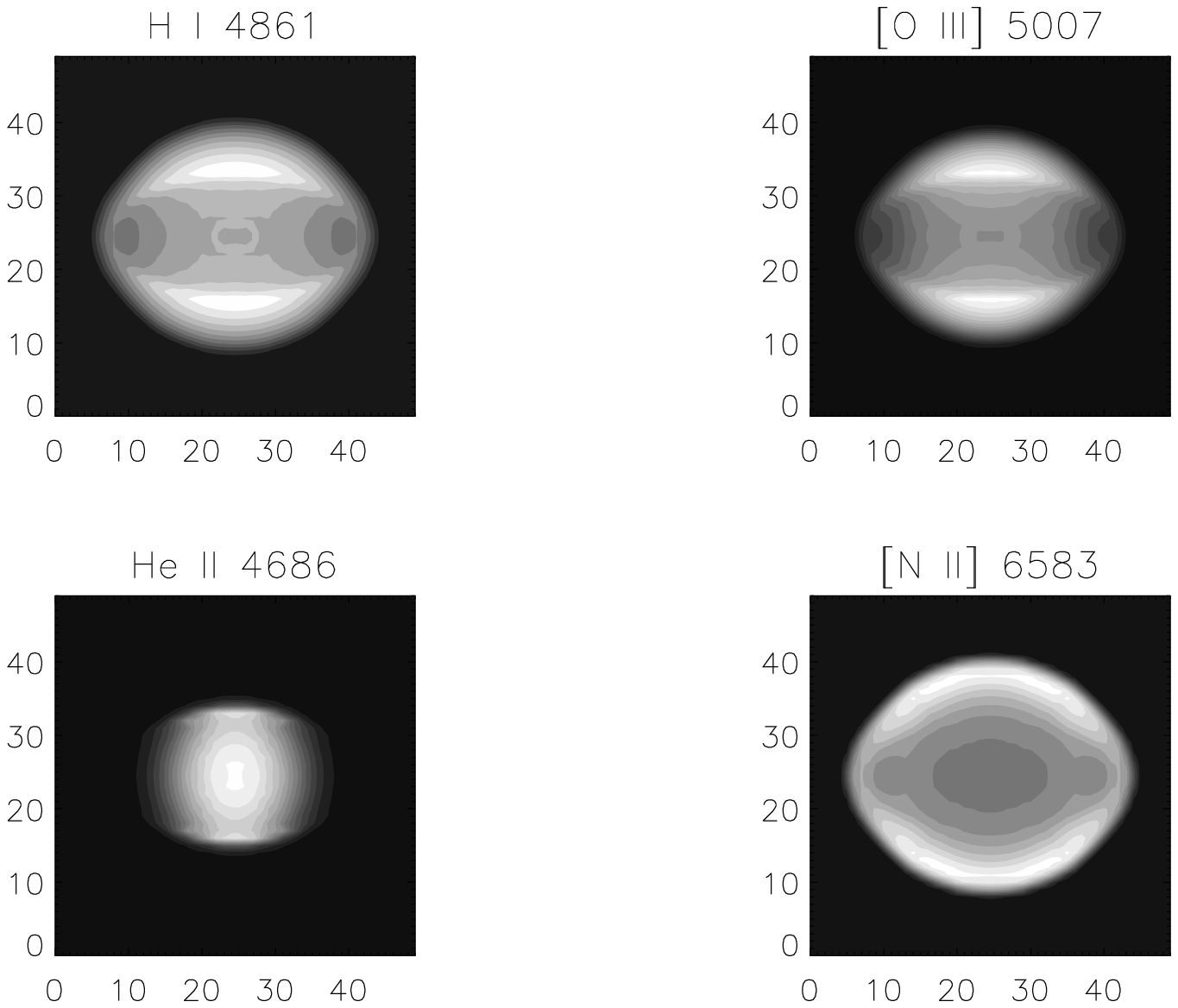]{
Same as Fig. \ref{fig-AA} for models B1 (top four panels) and 
B2 (botton four panels). Both axes are in pixels.
\label{fig-BB}}

\figcaption[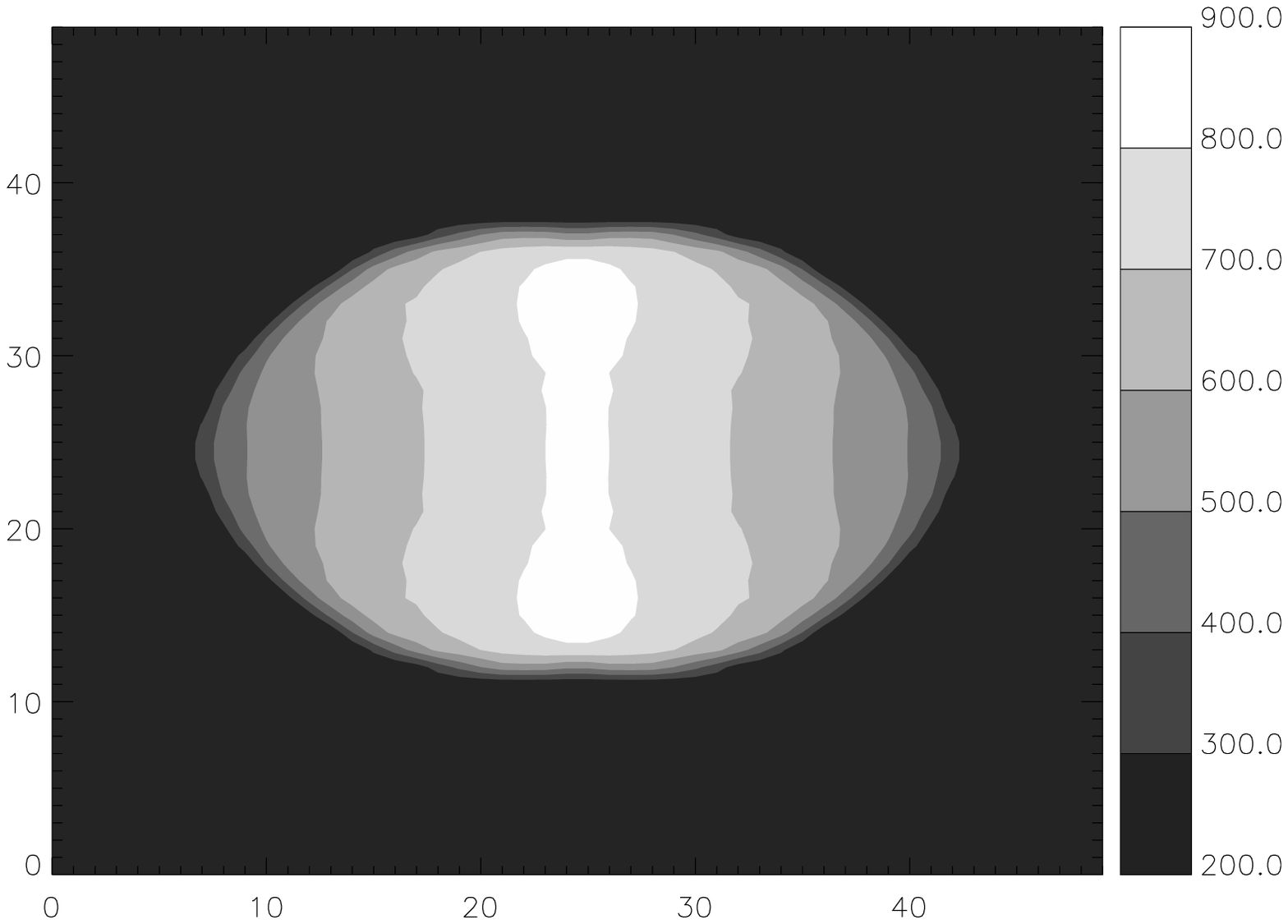,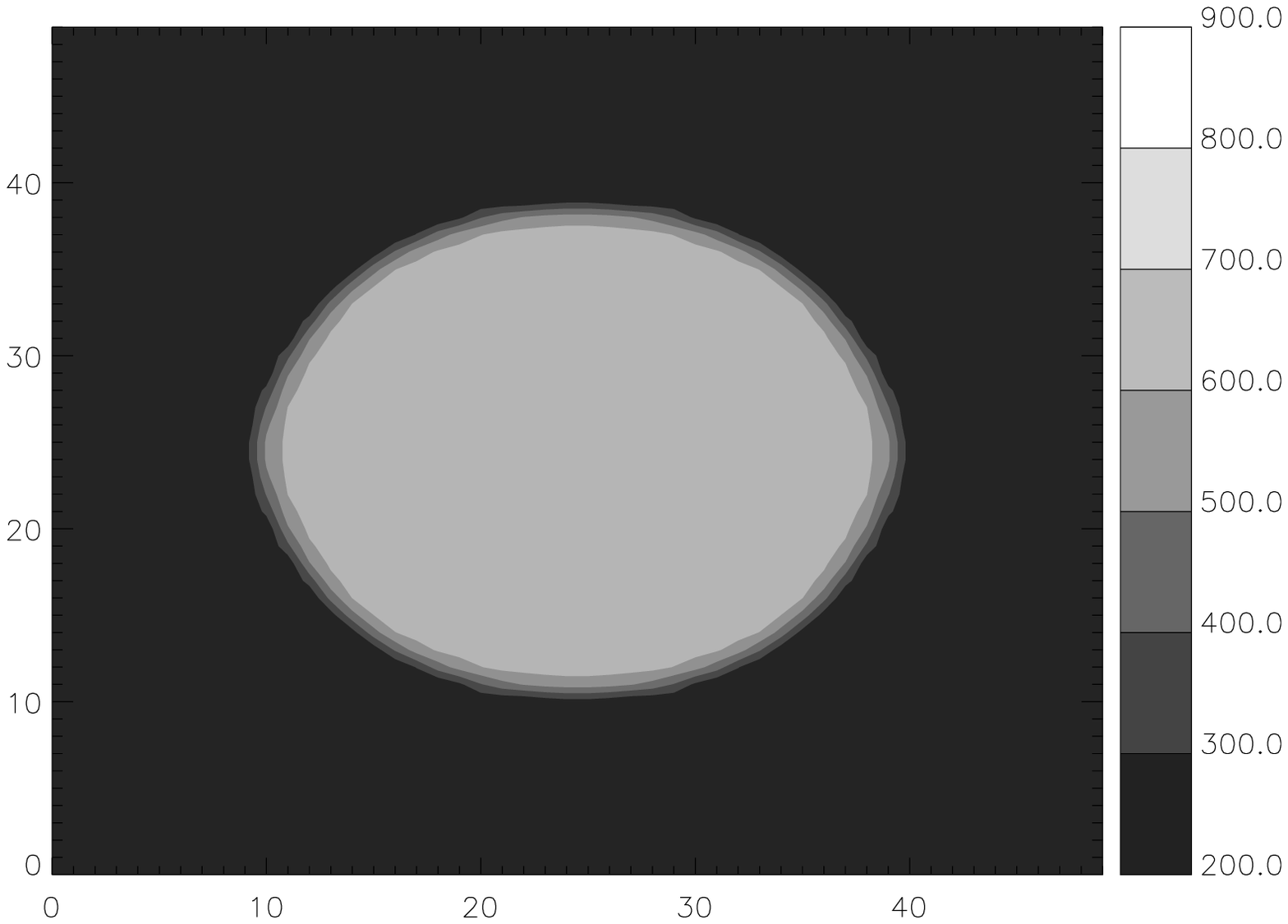]
{[S II] density map (from 200 to 900 cm$^{-3}$) for models B1 (top) and 
B2 (botton). The axes are in pixels.
\label{fig-rsii-A}}

\figcaption[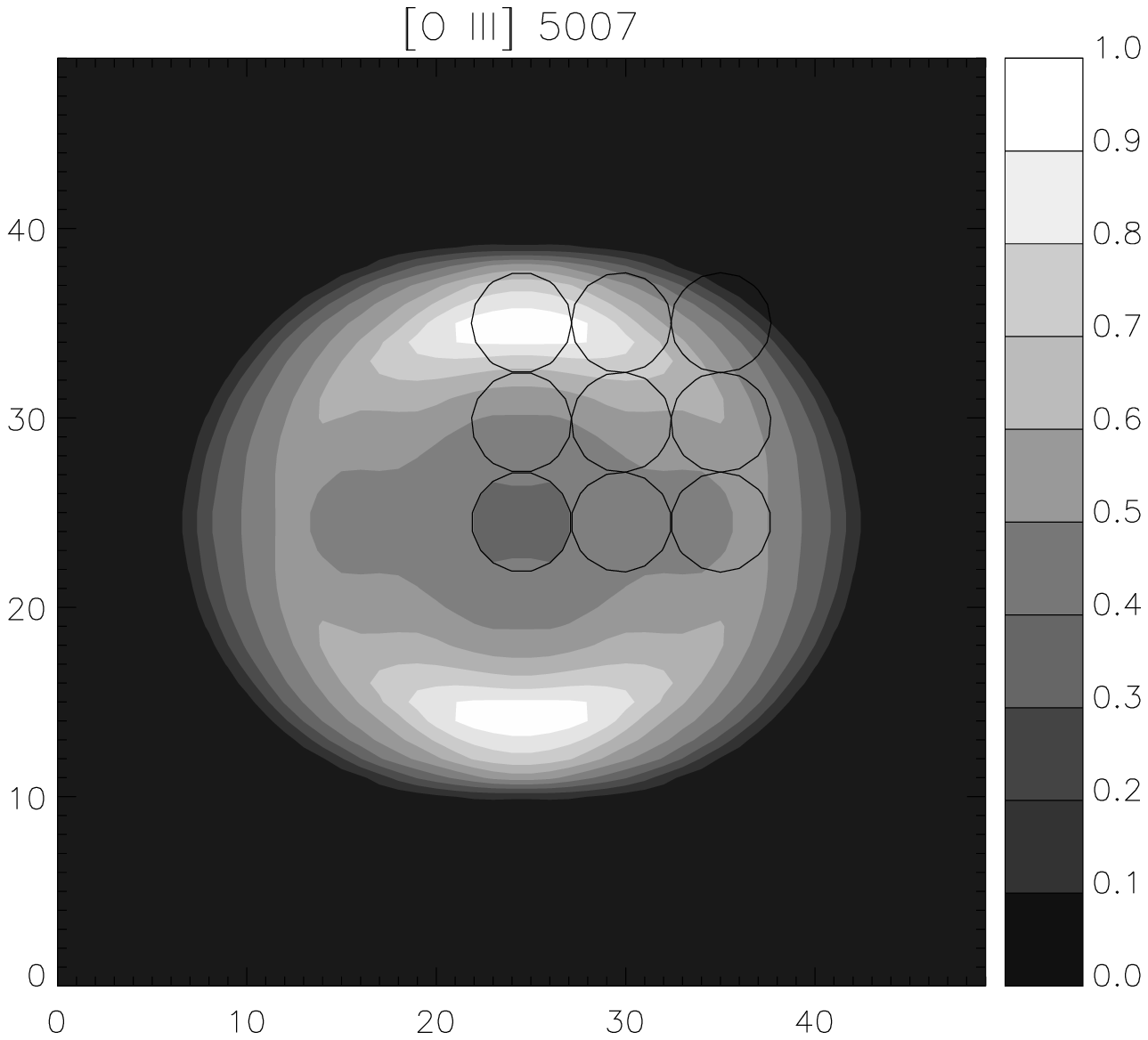,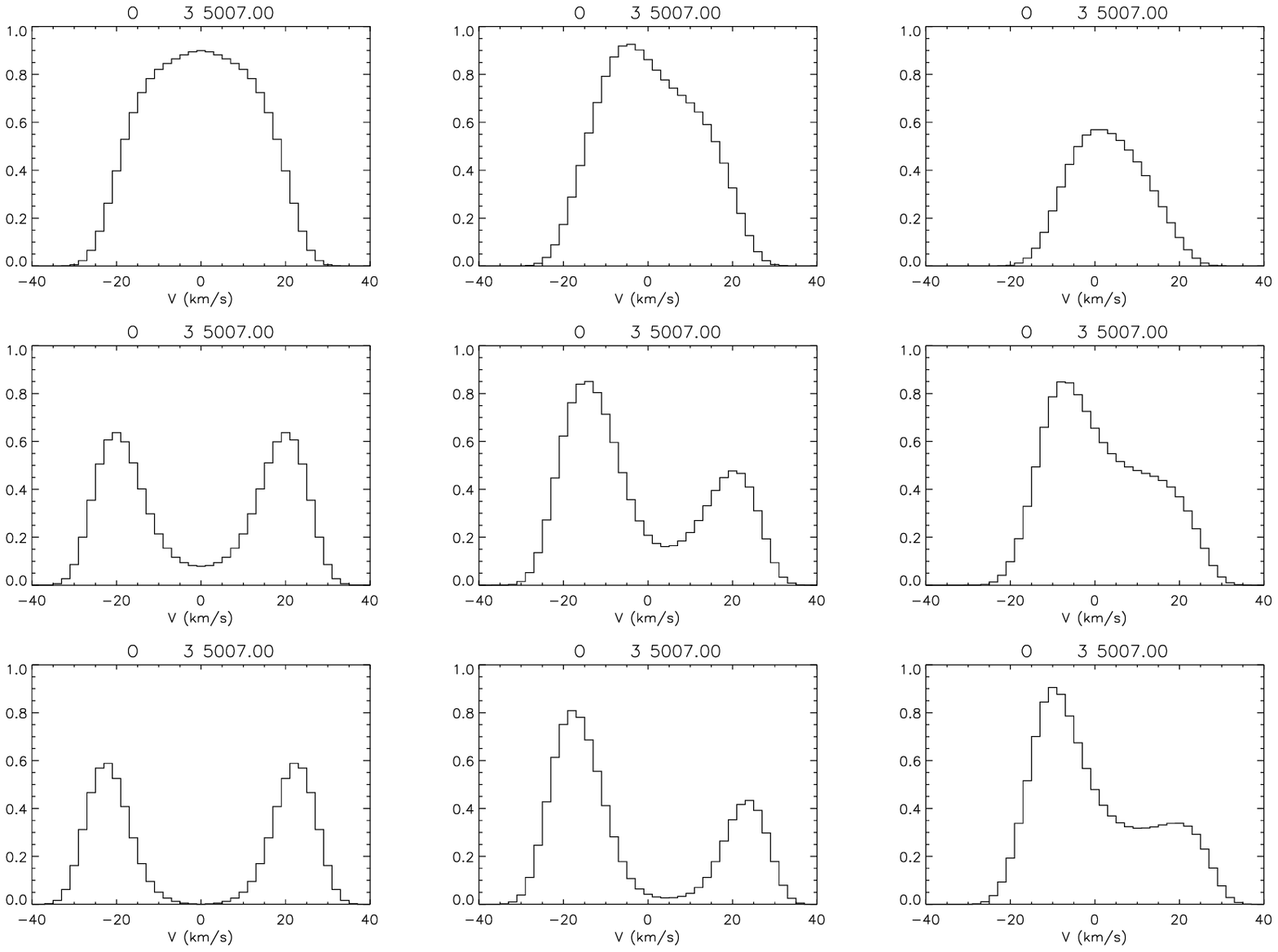]
{Size and positions of the aperture (top) used to obtain the
[O III] velocity profiles shown in the botton nine panels for model A1;
 The nebula is rotated by
50$^o$ around the $z$ axis and projected on the $y$ direction. The
axes are in pixels. The emission-line profiles (botton) are normalized.
\label{fig-velo_A}} 

\figcaption[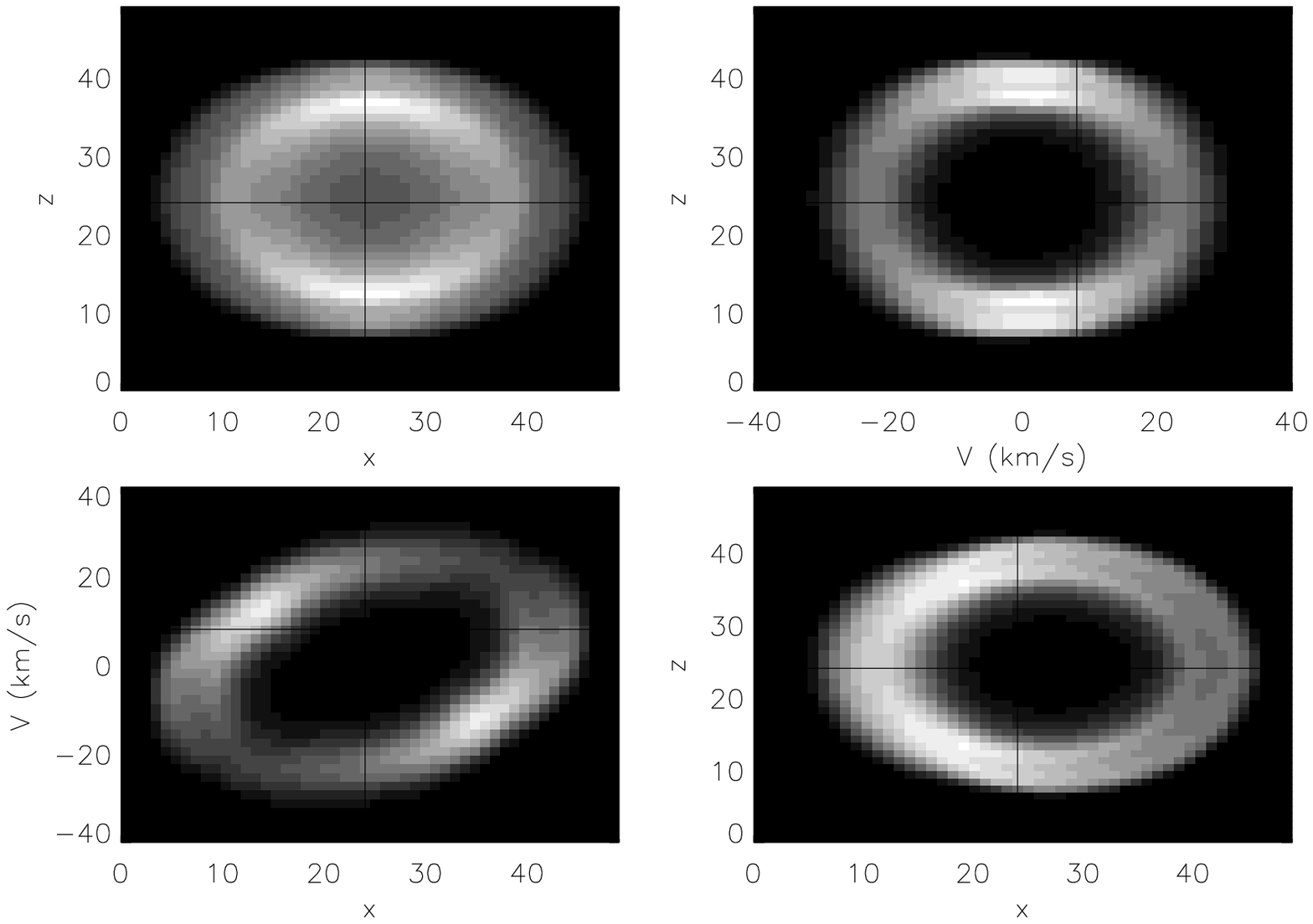]
{ [O III] line image (top left), P-V diagrams obtained through 
centered horizontal and vertical slits (top right and botton left,
respectively),  and iso-velocity contours for V = 9 km/s. $z$ and $y$
are given in pixels. 
\label{fig-ech_A}}

\newpage

\begin{table}[h]
    \begin{tabular}{lrrrr}
	\tableline
      Model 			& A1 	& A2	& B1 	& B2\cr 
	\tableline
      T$_{eff}$ (K) 		&150000	&150000	& 90000	& 90000 \cr
      L/L$_\odot$    		& 9260 	& 9260 	& 150 	& 150 \cr
      n$_H$(cav) (cm$^{-3}$) 	& 0 	& 0 	& 400 	& 400 \cr
      n$_H$(shell) (cm$^{-3}$)  & 2000-4000 & 3000 & 550 - 1100 & 825 \cr
      R$_{inner}/10^{17}$cm 	& 1. 	&0.5-1.5&1. 	& 0.5 - 1.5 \cr
      \tableline
      H$\beta$ (10$^{35}$erg/s) & 1.89  & 1.88 	& .048	& .047 \cr
      HeII4686 			& .330	& .328 	& .065	& .065\cr
      [OI]6300+6363		& .227	& .232 	& .402	& .392\cr
      [OII]3727 		& 1.83	& 1.97 	& 2.72	& 2.63 \cr
      [OIII]5007+4959  		& 24.0 	& 23.6 	& 6.72	& 6.92\cr
      [OIII]4363 		& .202  & .197 	& .018	& .019\cr
      [NII]6583+6548 		& 1.88 	& 1.90 	& 3.08	& 2.99\cr
      [NII]5755 		& .032 	& .032 	& .034	& .033\cr
      [SII]6718 		& .201	& .213 	& .608	& .598\cr
      [SII]6732 		& .297	& .307 	& .624	& .607\cr
      \tableline
      T$_{[OIII]}$ (K) 		& 12173	& 12122	& 8558	& 8567\cr
      $<T_e>_{5007}$ (K)	& 12021	& 11978	& 8591	& 8596\cr   
      T$_{[NII]}$ (K)		& 11621	& 11565	& 9568	& 9586\cr
      $<T_e>_{6583}$ (K)	& 11194	& 11225	& 9582	& 9612 \cr 
      n$_{[SII]}$ (cm$^{-3}$) 	& 2313 	& 2083	& 657	& 629 \cr
      $<n_e>_{6718}$ (cm$^{-3}$)& 2491 	& 2272	& 655	& 620 \cr
      \tableline
    \end{tabular}

  \caption{Input parameters: effective
    temperature and luminosity of the central ionizing black body,
    cavity density, shell density, cavity radius or the semi-axis of 
    the elliptical cavity (top). 
    Emission line intensities relative to H$\beta$ (middle). 
    Physical quantities determined from line ratios
    and from average calculations, see \S \ref{sub-stat} (bottom).
	 \label{tab-spectro}}
\end{table}

\newpage
\setcounter{figure}{0}

\begin{figure*} 
\plotone{fig1.1.ps} 
\plotone{fig1.2.ps} 
\caption{}
\end{figure*} 

\begin{figure*} 
\plotone{fig2.1.ps} 
\plotone{fig2.2.ps} 
\caption{}
\end{figure*} 

\begin{figure*} 
\plotone{fig3.1.ps} 
\plotone{fig3.2.ps} 
\caption{}
\end{figure*} 

\begin{figure*} 
\plotone{fig4.1.ps} 
\plotone{fig4.2.ps} 
\caption{}
\end{figure*}

\begin{figure*} 
\plotone{fig5.ps} 
\caption{}
\end{figure*} 

\end{document}